Enhancing the Adhesion of Graphene to Polymer Substrates by Controlled Defect Formation

*George Anagnostopoulos[1,*], Labrini Sygellou[1,*], George Paterakis[1,2], Ioannis Polyzos[1], Christos A. Aggelopoulos[1] and Costas Galiotis[1,2]*

[1]Institute of Chemical Engineering Sciences, Foundation for Research and Technology – Hellas (FORTH/ ICE-HT), Patras 265 04, Greece

[2]Department of Chemical Engineering, University of Patras, Patras 26504, Greece

Corresponding Author

Whom all correspondence should be sent to: Dr. Labrini Sygellou (sygellou@iceht.forth.gr) and Dr. George Anagnostopoulos (george.anagnostopoulos@iceht.forth.gr)

**ABSTRACT**

The mechanical integrity of composite materials depends primarily on the interface strength and the defect density of the reinforcement which is the provider of enhanced strength and stiffness. In the case of graphene/ polymer nanocomposites which are characterized by an extremely large interface region, any defects in the inclusion (such as folds, cracks, holes etc.) will have a detrimental effect to the internal strain distribution and the resulting mechanical performance. This conventional wisdom, however, can be challenged if the defect size is reduced beyond the critical size for crack formation to the level of atomic vacancies. In that case, there should be no practical effect on crack propagation and depending on the nature of the vacancies the interface strength may be in fact increase.

In this work we employed argon ion ($Ar^+$) bombardment and subsequent exposure to hydrogen ($H_2$) to induce (as revealed by X-ray & Ultraviolet photoelectron spectroscopy (XPS/UPS) and Raman spectroscopy) passivated atomic single vacancies to CVD graphene. The modified graphene was subsequently transferred to PMMA bars and the morphology, wettability and the interface adhesion of the CVD graphene/PMMA system were investigated with Atomic Force Microscopy technique and Raman analysis. The results obtained showed clearly an overall improved mechanical behavior of graphene/polymer interface, since an increase as well a more uniform shift distribution with strain is observed. This paves the way for interface engineering in graphene/polymer systems which, in pristine condition, suffer from premature graphene slippage and subsequent failure.

**Keywords:** CVD graphene /PMMA, $Ar^+$ bombardment, strain distribution, defects, adhesion force

**Introduction**

Graphene, an emerging nano-carbon material of atomic thickness[1] shows considerable promise in structural composite applications thanks to its unique combination of high tensile strength, Young's modulus [2] and structural flexibility [3, 4], which arise due to its structural (2D) perfection and its high strength and stiffness of the covalent C-C bonds [5]. Furthermore, thin sheets of graphene are amenable to new textural modifications such as those inspired by Japanese Kirigami[6], which could be stretched up to 240% of their initial length and bent using the radiation pressure of a laser beam or twisted with a magnetic field without breaking.

The excellent mechanical [7, 8], electrical [9, 10], as well as, thermal[11] properties of graphene can be put to good use when graphene is combined with other materials such as engineering plastics. In most semi-industrial scale applications [12-14] graphene is incorporated in the form of exfoliated flakes of μm dimensions but the obtained results are often limited by the small lateral size and ineffective dispersion [15, 16]. One way to tackle these problems particularly in certain applications, is to employ CVD-grown graphene [17, 18] that could be produced in either large or even continuous (roll-to-roll) sheets [12, 14] that can be incorporated on the top of polymer substrates or even embedded into polymer matrices [19].

However, in this case the graphene sheets contain structural imperfections such as a network of wrinkles, cracks and folds [20-25], which can generally be defined as defects [26-30] and impair somewhat the properties of the finished products. According to Han et al. [31], the onset of crack nucleation occurs near the presence of defects, determine graphene's performance under tensile loading. Moreover, the low interfacial adhesion between graphene and polymer due to the weak van der Waals bonding is still a problem that can lead to premature failure of CVD graphene/polymer systems at small strains [32]. Nevertheless, deviations from the perfect atomic arrangements in graphene play an important role in affecting its performance [19, 33-35], as they make it possible to tailor its local properties in cases where it is applied as a multi-functional coating [3, 36] in polymer matrices [37-40] for various composite applications [12, 14, 41-44] and to achieve new functionalities [45, 46].

The most frequently employed polymer is the poly-(methyl methacrylate) (PMMA), which is used as a substrate for graphene transfer and as a matrix for device fabrication. Compared to other typical polar molecules, such as polycarbonate or, even water, PMMA is a long-chain molecule with strong dipole interaction with graphene (attractive inter-surface forces such as van der Waals forces). Density Functional Theory (DFT) calculations have shown that PMMA radicals may form covalent bonds with a graphene defects, which usually appeared in grain boundaries. In particular, the local re-hybridization of carbons causes $sp^2$ to $sp^3$ transitions; thus, modifying graphene's band structure near the Fermi level [47].

Several strategies are proposed to modify the graphene surface in order to strengthen the interfacial adhesion [48-50]. Generally, irradiation by electron beam [51], or by ultra-violet (UV) radiation [52] or by ion radiation at room temperature gives rise to a continuous formation of defects and leading eventually to the development of holes and amorphization.

Among them, argon (Ar$^+$) ion bombardment is the most studied method and the defects are quantified mainly with Raman spectroscopy [53-55]. It has been found that when a population of defects is generated in a graphene sheet by irradiation and then is exposed to hydrogen, dissociative adsorption of hydrogen molecules is possible to occur [56]; thus, a better interfacial adhesion [45, 46] can be attained. In this context, it has been suggested that this dissociative adsorption of hydrogen molecules occurs on a graphite surface at vacancies [57] and at the armchair edge of graphite [58].

Based on the above a method to improve the interfacial adhesion between CVD graphene/PMMA is proposed herein. The main idea is to create a uniform network of atomic vacancies beyond the critical size for any crack or dislocation propagation. These vacancies can be treated (passivated) by chemical means in such a way that can act as anchors between graphene and surrounding polymer and thus enhance the interfacial adhesion. As presented below, the defects are created in an Ultra High Vacuum chamber (UHV) by Ar$^+$ bombardment on a CVD graphene grown on copper (Cu) substrate and the defected samples are exposed to $H_2$.

Such a procedure leads to a high density of defects on graphene surface [54] which can be identified by a variety of analytical techniques such as X-ray photoelectron spectroscopy (XPS). Furthermore, contact angle measurements showed that this procedure also leads to more hydrophobic surfaces, while Raman analysis along with atomic force microscopy (AFM) topography are employed for assessing the strain distribution in the graphene.

## 1. Experimental

*Synthesis of CVD graphene grown on Cu:* CVD synthesis of graphene was performed in an AIXTRON® (Black Magic) CVD chamber. Copper foils supplied from JX Nippon Mining & Metals® with a purity of 99.95% were used as the catalyst substrates. Before the introduction of the copper foil into the CVD chamber, the foil was cut into 7 x 7 cm$^2$ and cleaned by isopropanol to remove any organic contamination. After the closure of the chamber, the foil was heated in 1000 °C in argon/hydrogen atmosphere and was kept there for 5 min for annealing. Then, the hydrogen flow was terminated, temperature was decreased until 925 °C and at the same time methane was introduced into chamber, as carbon feedstock to initiate the graphene growth on copper foil surface. After 5 min at 925 °C, the chamber was cooled down to 650 °C, where the methane flow was terminated. Finally, the chamber was cooled gradually to room temperature.

*Ar$^+$ ion bombardment:* The CVD graphene/Cu samples were introduced in UHV chamber and subjected to Ar$^+$ bombardment without previous treatment in order to avoid changing its pristine properties. The only parameter which was varied was the ion energy, from 35 up to 200 eV, while the Ar$^+$ pressure, the irradiation time and the spot were constant. The Ar$^+$ pressure was $3 \times 10^{-6}$ mbar, the duration 12 seconds, the distance between the sputtering gun and the sample was 6 cm while the spot size was a circle with 7 mm diameter. Subsequently after irradiation, the samples were exposed to hydrogen atmosphere (1bar) for 10 min in the high pressure chamber.

*Samples' preparation and transfer to polymer substrates:* The modified graphene surfaces were transferred on a PMMA substrate. Both pristine and treated CVD graphene/Cu samples were transferred from the copper foil to a PMMA substrate, by implementing a *"dry transfer"* method. The steps of the method are shown schematically in Figure S2. With this method, graphene was actually transferred to the desired position, on the PMMA substrate, just before the removal of copper. Initially, a thin layer of PMMA ~200 nm (495k, 3% in anisole) was spin-coated on graphene and copper. After that, the graphene sample with the PMMA coated film and the PMMA substrate, were pressed for a couple of hours. Then, the uncoated graphene (back side of the copper foil) was removed by using $O_2$ plasma and the copper foil was etched by 0.15M ammonium persulfate (APS) solution. Finally, the sample was replenish with de-ionized water and dried under $N_2$ flow.

*Characterization by implementing XPS/UPS:* The surface analysis studies were performed in a UHV chamber (P<10mbar), which consists of a three chambers: (a) a high pressure chamber, where exposures of samples for gas pressure up to several atmospheres takes place, (b) a preparation chamber, where the ion gun for Ar sputtering is placed and (c) an analysis chamber. The analysis chamber is equipped with a SPECS LHS-10 hemispherical electron analyzer, a dual-anode x-ray source for XPS and a UV source (model UVS 10/35) for UPS measurements.

The XPS measurements were carried out at room temperature using unmonochromatized AlKa radiation under conditions optimized for maximum signal (constant $\Delta E$ mode with pass energy of 36 eV giving a full width at half maximum (FWHM) of 0.9 eV for the Au $4f_{7/2}$ peak). The analyzed area was an ellipse with dimensions 2.5 x 4.5 mm$^2$. The XPS core level spectra were analyzed using a fitting routine, which can decompose each spectrum into individual mixed Gaussian-Lorentzian peaks after a Shirley background subtraction. The UPS spectra were obtained using HeI irradiation with hv = 21.23 eV produced by a UV source (model UVS 10/35). During UPS measurements the analyzer was working at the Constant Retarding Ratio (CRR) mode, with CRR = 10. The work function was determined from the UPS spectra by subtracting their width (i.e. the energy difference between the analyzer Fermi level and the high binding energy cutoff), from the HeI excitation energy. For these measurements a bias of −12.30 V was applied to the sample in order to avoid interference of the spectrometer threshold in the UPS spectra.

*Raman measurements:* In order to check the influence of the morphology of the substrate on graphene properties, Raman mapping took place. Spectra were taken with at 514 nm (2.41 eV) laser using a MicroRaman (InVia Reflex, Rensihaw, UK) set-up. The laser power was kept below 1.5 mW on the sample to avoid laser-induced local heating, while an Olympus MPLN100x objective (NA = 0.90) was used to focus the beam on the samples. The obtained spectra of the pristine and treated CVD/PMMA system are presented in Figure S5.

*Contact Angle Measurements:* The contact angle of water drops on the surface of graphene/Cu samples were measured as follows: a droplet of the liquid was deposited by a syringe which was positioned above the sample surface and the image captured by a high resolution stereomicroscope (Nikon SMZ1000) was analyzed to determine the contact angles.

*Mechanical tensile tests:*

*Quasi-static deformation:* The top surface of the monolayer CVD graphene/PMMA system was subjected to tension using a four-point bending apparatus. A more detailed description of the mechanical testing applied is presented elsewhere [59]. In order to conduct Raman mapping during loading, the four-point bending apparatus was placed on a three-axis piezoelectric translation stage that was operated on three orthogonal axes by a ThorlabsInc. piezoelectric controller. The NanoMax three-axis flexure stage can provide nano-metric positioning on the three orthogonal axes. At each strain level, the stage was translated with a step of 5 μm with the simultaneous collection of Raman spectra within an area of 15 x 15 μm$^2$ (16 total points) for both examined systems.

*Atomic Force Microscopy measurements:* AFM images were collected by a contact mode (Bruker, Dimension-Icon). Images were obtained using ScanAsyst-Air probes (silicon tips on silicon nitride cantilever, Bruker) with 0.4 N m$^{-1}$ nominal spring constant of the cantilever.

## 2. Results and Discussion

### 2.1 Modification of the CVD graphene sheets

Since the monolayer graphene is only 0.3 nm thick [60, 61], surface sensitive methods, such as XPS, provide valuable information about the changes on graphene before and after irradiation. The defects on a CVD mono-layer graphene on Cu are implemented by $Ar^+$ ions irradiation in an UHV chamber. In order to avoid possible contamination of the defect sites of the irradiated surfaces during air exposure, the irradiated samples were subsequently exposed to $H_2$ atmosphere.

It has been shown that hot hydrogen atoms (H) (i.e. with a few tens of eV) can be adsorbed on the basal plane of surface-clean graphene, while adsorption is barrier-less on free edges or vacancies (dangling bonds) [62]. In the present study, the effect of $H_2$ exposure on the $Ar^+$ irradiated CVD graphene/Cu is investigated and it was found that H adsorbed on the defect sites created by $Ar^+$ irradiation, prevented oxygen contamination from air [63]. The details are presented in the Supporting Information.

Subsequently, a method to induce defects with a controlled density is developed, by adjusting the energy of the $Ar^+$ that interact with the surface. The ion energy was changed by regulating the voltage of the anode in the ion gun, while the irradiation time remained constant (12s). The number of defects per $cm^2$ of the graphene area was calculated by measuring the surface current during $Ar^+$ irradiation and assuming that each $Ar^+$ ion interacts with one carbon atom. The corresponding results for each ion energy are shown in Table 1 [64].

After each $Ar^+$ energy irradiation and $H_2$ exposure in-situ XPS-UPS and ex-situ, Raman measurements recorded. Figure 1 shows the deconvoluted C1s XP spectra for pristine (Figure 1a), $Ar^+$ irradiated and $H_2$-exposed CVD graphene/Cu samples for different $Ar^+$ energies. In order to remove the background noise caused by inelastic electron scattering, a simple Shirley-type correction was introduced.

The C1s peaks are analyzed into four components at binding energies $284.70 \pm 0.05$ eV assigned to the C=C $sp^2$ bonds, at $285.60 \pm 0.05$ eV attributed to carbon atoms with $sp^3$ hybridization and two components and at 286.70 ± 0.1 eV and 288.5 ±0 .1 eV assigned to C-OH and O-C=O bonds, respectively [65]. The Ar$^+$ impact per cm$^2$ area has been directly correlated to the $sp^3/sp^2$ ratio and the results are shown in Table 1. The $sp^3/sp^2$ intensity components ratio for the pristine sample is 0.10 corresponding to ~9% $sp^3$, which comes into agreement with the results obtained by Luo et al. [66]

Figure 2 shows the HeI UPS valence band spectra and the high binding energy cutoff, from the pristine and the treated CVD graphene/Cu surfaces, respectively. In Figure 2I, the region 0-12 eV consists mainly of 3 peaks:(i) at ~3 eV (labeled as A) attributed to pπ electrons (sp$^2$ hybridization), (ii) at ~6 eV (labeled as B) assigned to $2p\pi+\sigma$ and (iii) at~ 10 eV (labeled as C) assigned to $2s$-$2p$ hybridization[67]. No great differences seems to be present in the spectra for Ar$^+$ energies of range 35-130 eV, whereas at 200 eV the peak at ~6 eV is decreased in comparison to the peak at ~10 eV, while a reduction of the density of $2p\pi$ states (~3 eV) near the Fermi level is observed. These changes in the valence band are expected since the $sp^3$ C-H defects in graphene, depresses the delocalized π-electrons significantly.

Figure 2II shows the cut-off of the high binding energy region, where the work function (WF) of graphene can be estimated by subtracting the width of the photoelectron spectrum from the photon energy. The WF of the pristine surface is 4.40 ± 0.05 eV in agreement with literature values for a pristine graphene [67]. For the irradiated surfaces, the WF is gradually decreased starting from 65 eV Ar$^+$ energy irradiation and reaching the value of 4.20 ± 0.05 eV for 200 eV Ar$^+$ energy (Δφ=0.2eV).

It is known that the $sp^3$bonding of carbon surface terminated in hydrogen leads to a small decrease of the work function which arises from the formation of a C$^{-\delta}$–H$^{+\delta}$ surface dipole layer and its magnitude is proportional to the surface coverage of hydrogen [68]. Thus, the slight decrease of the WF of the treated surfaces is an indication of low coverage by H of the defected sites.

The Raman spectra of the pristine graphene/Cu and the treated surfaces are showed in Figure S2 in the supplementary. The Raman spectrum of graphene has the typical peaks of all carbon allotropes[69]. Briefly, the G and D peaks, around 1580 and 1360 cm$^{-1}$ are due to the s$p^2$bonded carbons. The G peak corresponds to the $E_{2g}$ phonon at the Brillouin zone (BZ) center ($\Gamma$ point). The D peak is due to the breathing modes of six-atom rings and requires a defect for its activation. It comes from TO phonons around the BZ K point, and it is activated by an intra valley scattering process. The 2D peak is the second order of the D peak. The defect activated D´ peak comes from the LO branch of the phonon dispersion and is seen around 1620 cm$^{-1}$.

The evolution of Raman spectrum follows the trend firstly reported by Cançado et al.[54]; thus, the increase of bombardment dose is followed by the appearance and enhancement of the D peak (35-65 eV) and the subsequent increase of the ratio I(D)/I(G). As the applied ion-dose further increases, the D´peak rises and all Raman peaks broaden (~120 eV). Due to broadening, the G and D´ peaks tend to overlap and form a single wider and blue shifted peak, while a sharp decrease of the ratio I(D)/I(G) is observed (130-200 eV). Therefore, the decrease of the ratio I(D)/I(G) indicates here enhancement of disorder since the G peak incorporates in essence the disordered-induces D´ peak. The cut-off dose for defect creation is ~35 eV, whereas above ~200 eV graphene is severely damaged.

Based on the obtained Raman spectra (Figure S2a,b) and the dependence of the ratio I(D)/I(G) versus the bombardment dose applied (Figure S2c), the energy of 120 eV is selected (optimum energy applied), as a trade-off between the number of defected areas and their size [54], in order to further investigate the stress transfer efficiency of defected graphene/PMMA system.

*2.2 Evaluating the morphology and the interface integrity of the CVD graphene coating/PMMA system before and after the insertion of defects*

The morphology and the interface integrity of the pristine and the treated CVD graphene/PMMA system was evaluated for several samples. Initially, AFM was employed for both systems, to assess the structure of the as supported CVD graphene at the nanoscale. As can be seen from AFM high resolution images (Figure 3), two different morphologies appeared depending on whether or not they had been treated by Ar[+] irradiation. The first region (Figure 3a, c) is relatively "*flat*" while the second that has a "*rugged*" appearance corresponds to the presence of folds without a preferred orientation (Figure 3b, d).

These areas seem to be created on Cu foil, at the cooling step of CVD process [70, 71] and then are transferred directly on PMMA bar during the transfer process (Figure S3) (see also experimental section). The creation of folds presented in Figures 3b and 3d are very similar to the folds appeared on graphene oxide films due to applied biaxial compression [72]. The latter seems to be confirmed also for the samples examined by Raman measurements (Table S1), as it will be discussed further in the text. Similar images obtained from other specimens are also presented in Figure S5.

In addition to the topographical study, adhesion force measurements were also recorded. The obtained statistical analysis (histograms) of the adhesion forces show an unsymmetrical Gaussian distribution for the pristine (both for the "*flat*" and "*rugged*" areas) and more symmetrical for the treated graphene, respectively (Figure 3a(ii) and 3c(ii)). Similarly, the maximum value of the Gaussian distributions is shifted from 3.22 nN at the pristine state to higher force values of 5.18 nN for the treated surface, respectively. Analogous behavior was observed for the "*rugged*" regions (Fig. 3b(ii) & 3d(ii)), where the mean value of the adhesion force is increased from 2.85 nN for the pristine to 5.84 nN for the treated specimens, respectively.

Based on the analysis made by Jiang et al. [73] (see also Supporting Information), the adhesion energy between graphene and PMMA substrate can be measured by force spectroscopy mode of AFM, using the Maugis-Dugdale model [74]. As depicted in Table 2, there is a great increase of the surface energy, reaching up to 115% and 66% for the "*rugged*" and "*flat*" regions, respectively. As it is stated Jiang et al. [73], the surface roughness significantly affects the measured adhesion force (Table 2). By introducing a controllable amount of defects, the graphene's roughness; thus, affecting further its interaction with the PMMA substrate.

Such an increase of the adhesion force can only be attributed to the modification of the graphene surface by the insertion of defects followed by the $H_2$ exposure. It seems that such a functionalization improves the bonding between graphene and hydrophobic PMMA transfer film (see also Experimental and Figure S3), as it is also confirmed by the macroscopic results obtained from the contact angle measurements (Figure S4), where an increase of 10% is recorded (Table 3).

The effect of the presence of defects in the wettability of graphene has been studied for oxygen plasma treated graphene on SiC [75]. As the density of defects induced by plasma treatment increases, the surface energy also increases due to the fact that the graphene become polar after creating polar O−H bonds leading to a hydrophilic nature. In our case, the passivation of treated graphene with hydrogen atoms results just the opposite. As it is stated by Wu et al. [63], during the preparation of graphene sheets by thermal exfoliation of graphene oxide (GO), the presence of hydrogen is essential for de-oxygenation of GO as well for structural defects removal; thus, oxygen-containing groups, i.e. hydrophilic surface functionalities, lead to a formed graphene with superior properties.

By following Neumann's model [76] (see also Supporting Information) and by applying a water droplet (polar solvent) before and after the treatment, the obtained surface energy of graphene on Cu is reduced by 5.9% (Table 3); thus, becoming less hydrophilic. Therefore, the changes observed on the wetting behavior of graphene seem to affect positively its interaction with the non-polar PMMA film during the transferring process; thus, better interactions with the PMMA substrate (bar), leading to different mechanical response to the external applied load, as it will be further discussed below.

Additionally, by analyzing the profiles of the most representative Raman peaks of graphene (2D, G and D) [77-79] for both graphene systems, it seems to lead to the same result. By carefully selecting the same sampling areas of 20 x 20 $\mu m^2$ presented in Figure 4, a full Raman investigation (121 data points) from both systems are obtained (Figure S6a and b). The statistical analysis of the above is presented in Figure S7.

For both systems, Pos(2D) has a mean value of the order of ~ 2690 cm$^{-1}$, indicating the presence of compression and/or doping[59] (Figure S7a, b). If we reasonably consider that the Pos(2D) shift is due to the imposition of biaxial strain during the production process[80], the corresponding compressive biaxial strain calculated, using the sensitivity value +148 cm$^{-1}$%[80], varies between -0.07% and -0.06% for the pristine and treated system, respectively (Table S3).

By correlating Pos(2D) versus Pos(G) as shown in Figure S8, it can be argued that for the collected data points the mechanical loading dominates upon doping[77], since the majority of the points follow a linear relationship. Especially, for the case of the treated CVD graphene the least-squares fitted slope (~1.7) is greater than the corresponding for the pristine system (~1.4), implying a greater mechanical adhesion [59, 81].

As for the Raman linewidths, useful interpretations can be extracted regarding doping, strain, disorder and number of layers in graphene[82, 83]. For the pristine CVD synthesized graphene film, the corresponding value of FWHM(2D) (33.3±4.2 cm$^{-1}$) (Figure S7b) at rest is larger compared to exfoliated flakes (~24 cm$^{-1}$) [84] and the reasons for that are given elsewhere [84, 85]. However, there is a set of data points exhibiting values of FWHM(2D) greater than 33 cm$^{-1}$, corresponding to ~20% of the examined data points (Figure S7b), which is attributed to bi-layer or even multilayer islands (three layers or more) [86, 87], as it has been explained elsewhere [77, 86-90]. For the treated CVD synthesized graphene sheet, the increased value of FWHM(2D) (Figure S7d) occurs as a result of the defect insertion, as it is also confirmed in Figure S2a. Similar results are observed for the FWHM of the G peak (Figure S7g and h)

In addition, the presence of defect-activated D band[54, 91] is relatively lower for the pristine case, as it is confirmed by the value of ~0.50 for the I(D)/I(G) ratio (Figure S7i). The latter indicates that the pristine examined system has indeed a small amount of disorder or defects [92, 93]. According to Cançado et al.[54, 94], by the insertion of defects, (Figure S2 and S6b) the intensity ratio of the D and G Raman peaks (I(D)/I(G)) increases (Figure S7i and j), the D′ peak appears (Figure S6b), and a broadening of all peaks is also observed (Figure S2 and S6b)[91]. By implementing a controllable defect insertion, the obtained corresponding value is of the order of ~3 (Figure S7j).

Following the analysis of Eckmann et al. [55], for the relative intensity of the D and D′ peaks (Figure S9), it is suggested that the inserted defects, for the optimum energy 120eV (Figure S2c), are actually single-vacancy defects, since the value of I(D)/I(D′) is of the order of ~8.5. Similar results were also obtained by Polin et al.[95], who argued that graphene becomes stiffer by controlled defect creation. As it will be shown below, it seems that those vacancies are actually assisting graphene to interact better with the polymer substrate (PMMA bar), as it is proved by the increase of the adhesion force (Figure 3).

### *2.3 Implementation of mechanical loading*

The mechanical response for both pristine and treated CVD graphene /PMMA systems upon external deformation (up to 1.0%) for an area in the specimens of 15 x15 $\mu m^2$ was investigated. Prior to loading the strain state of the examined specimens was identified within the allocated sampling areas and the residual compressive strain is presented in Table S3.

An important parameter that has been identified by us and others previously [37, 96-98] as an index of stress-transfer efficiency, is the Raman shift rate per strain for the two most common mentioned vibrational modes (2D and G). For exfoliated graphene/ PMMA systems the maximum values recorded are in the range of 55-60 $cm^{-1}$/% for the 2D peak and ~25$cm^{-1}$/% for the mean value of the G peak [99]. For corresponding CVD graphene/ PMMA systems the above values represent upper limits often difficult to attain due to the inherent morphological defects of the CVD-grown graphene [19, 34, 100]. In this context, strain rate maps extracted for both loading and unloading cycles can actually be considered as an indirect adhesion indicators, since the measured Raman variations shifts of Pos(2D) and Pos(G) are related with the "*true*" strain transferred from the matrix to the inclusion.

During loading, a broad value range for the strain rate of Pos(2D) is observed for the pristine CVD graphene /PMMA system (Figure 4a). Particularly, there is a group of points (Group A, Figure 5a), for which the Pos(2D) is shifted to lower values at very low rates (from -5 to -15 $cm^{-1}$/%).

However, there are other points (Group B), for which the strain rates reach -55 cm$^{-1}$/% [101], which is the upper limit as mentioned above and also confirmed by others [37, 97-99].

Similarly, the Pos(G) red-shifts at a rate of -6 to -3 cm$^{-1}$/% for Group A, while the points of Group B exhibit shits of up to -17 cm$^{-1}$/% (Figure S10a). Moreover, for Group B, splitting of both the G and 2D peaks are observed which are indicative of efficient loading of the inclusion [102, 103]. The latter is depicted in Figure S11, where typical spectra from both point groups are presented for applied strain of 0.80%. During unloading, both groups of points are shifting back at similar shift rates as the loading curves indicating elastic behavior (Figure 4b, Figure S10b).

On the contrary the treated CVD graphene/PMMA system exhibits on average a much higher and narrower distribution of Pos(2D) shift rates during loading (Figure 6a), of ~-30 cm$^{-1}$/%. Therefore, it can be stated that the overall interaction with the substrate is higher than to the cases prior to treatment and therefore the stress transfer efficiency has been greatly improved. Similar results are obtained for the Pos(G) which shifts approximately by ~-10 cm$^{-1}$/% on average (Figure S12a), very close to the value of -14.7 cm$^{-1}$/% obtained elsewhere [37] for CVD graphene. During unloading (Figure 5b, S12b), all the points of the examined area are blue-shifting with almost the same strain rate for both bands.

Figure 6 shows the section analysis of AFM images prior and after loading for the pristine and treated CVD graphene, respectively. After loading, it seems that the treated graphene has returned to a great extent to its initial loading position, compared to the pristine specimen (Figure S13). Furthermore, the statistical analysis of the relative change of surface's height ($\Delta H/H_o$) prior and after loading, shows an unsymmetrical Gaussian distribution for the pristine and more symmetrical for the treated graphene, respectively (Figure 6a(iii) and 6b(iii)).

For the former, recent works suggests that the repetitive reforming and breaking of interaction at interface region would occur during sliding process for both van der Waals and H-bonds interactions [104, 105]. In case of the treated system, the insertion of specific population of defects on the CVD graphene along with their covering by H atoms, affects its interaction with the PMMA surface to a greater and more homogeneous extent; thus, graphene interacts better with the polymer substrate caused by a strong interaction of the defects with the PMMA.

### 3. Conclusions

In this work, CVD graphene on Cu substrate is treated by inserting defects with a well-controlled population by $Ar^+$ bombardment in an UHV chamber, followed by a subsequent exposure to hydrogen atmosphere. The treated surfaces were characterized by several techniques which revealed that hydrogen reacts with carbon at the defect sites and passivated the carbon atoms.

We have shown that a controlled generation of atomic defects to CVD graphene results in the moderate increase of the adhesion between graphene and a polymer substrate and leads to a more uniform strain uptake in reinforcing inclusion. Since graphene adheres to polymer substrates through weak van der Waals bonding, this approach paves the way for improving the mechanical behavior of graphene/ polymer interface and for eventually tailoring the mechanical properties of graphene. While the proposed method presented herein represent the interface of CVD graphene/PMMA, it can be applied to other 2D materials improving the interfacial mechanics of composite materials.

### ACKNOWLEDGEMENTS

The research leading to these results has received funding from research projects entitled: "*Graphene Core 1, GA: 696656 – Graphene-based disruptive technologies*" and "*Graphene Core 2, GA: 785219*", which are implemented under the EU-Horizon 2020 Research & Innovation Actions (RIA) and are financially supported by EC-financed parts of the Graphene Flagship. It is also acknowledged the support of the ERC Advanced Grant "*Tailor Graphene*" (no: 321124). In addition, it is acknowledged Dr. George Trakakis for the production of CVD graphene specimens on copper foil, Dr. Emmanuel N. Koukaras and PhD candidate Antonis Michail for the extraction of adhesion and surface energy values, as well PhD candidate Nick Koutroumanis for the flexural measurements of PMMA substrates.

**REFERENCES**

[1]  Novoselov, K. S.;Geim, A. K.;Morozov, S. V.;Jiang, D.;Zhang, Y.;Dubonos, S. V.;Grigorieva, I. V.; Firsov, A. A. Electric Field Effect in Atomically Thin Carbon Films. *Science* **2004**, *306*, 666-669.

[2]  Lee, C.;Wei, X.;Kysar, J. W.; Hone, J. Measurement of the Elastic Properties and Intrinsic Strength of Monolayer Graphene. *Science* **2008**, *321*, 385-388.

[3]  Jang, H.;Park, Y. J.;Chen, X.;Das, T.;Kim, M.-S.; Ahn, J.-H. Graphene-Based Flexible and Stretchable Electronics. *Adv Mater* **2016**, *28*, 4184-4202.

[4]  Sang Jin Kim, K. C., Bora Lee, Yuna Kim, and Byung Hee Hong Materials for Flexible, Stretchable Electronics: Graphene and 2D Materials. *Materials Research* **2015**, *45*, 63-84.

[5]  Fasolino, A.;Los, J. H.; Katsnelson, M. I. Intrinsic ripples in graphene. *Nat Mater* **2007**, *6*, 858-861.

[6]  Blees, M. K.;Barnard, A. W.;Rose, P. A.;Roberts, S. P.;McGill, K. L.;Huang, P. Y.;Ruyack, A. R.;Kevek, J. W.;Kobrin, B.;Muller, D. A.; McEuen, P. L. Graphene kirigami. *Nature* **2015**, *524*, 204-207.

[7]  Young, R. J.;Kinloch, I. A.;Gong, L.; Novoselov, K. S. The mechanics of graphene nanocomposites: A review. *Composites Science and Technology* **2012**, *72*, 1459-1476.

[8]  Vadukumpully, S.;Paul, J.;Mahanta, N.; Valiyaveettil, S. Flexible conductive graphene/poly(vinyl chloride) composite thin films with high mechanical strength and thermal stability. *Carbon* **2011**, *49*, 198-205.

[9]  Fiori, G.;Bonaccorso, F.;Iannaccone, G.;Palacios, T.;Neumaier, D.;Seabaugh, A.;Banerjee, S. K.; Colombo, L. Electronics based on two-dimensional materials. *Nat Nano* **2014**, *9*, 768-779.

[10] He, F.;Lau, S.;Chan, H. L.; Fan, J. High Dielectric Permittivity and Low Percolation Threshold in Nanocomposites Based on Poly(vinylidene fluoride) and Exfoliated Graphite Nanoplates. *Adv Mater* **2009**, *21*, 710-715.

[11] Teng, C.-C.;Ma, C.-C. M.;Lu, C.-H.;Yang, S.-Y.;Lee, S.-H.;Hsiao, M.-C.;Yen, M.-Y.;Chiou, K.-C.; Lee, T.-M. Thermal conductivity and structure of non-covalent functionalized graphene/epoxy composites. *Carbon* **2011**, *49*, 5107-5116.

[12] Bae, S.;Kim, H.;Lee, Y.;Xu, X.;Park, J.-S.;Zheng, Y.;Balakrishnan, J.;Lei, T.;Ri Kim, H.;Song, Y. I.;Kim, Y.-J.;Kim, K. S.;Ozyilmaz, B.;Ahn, J.-H.;Hong, B. H.; Iijima, S. Roll-to-roll production of 30-inch graphene films for transparent electrodes. *Nat Nano* **2010**, *5*, 574-578.

[13] Yamada, T.;Ishihara, M.; Hasegawa, M. Large area coating of graphene at low temperature using a roll-to-roll microwave plasma chemical vapor deposition. *Thin Solid Films* **2013**, *532*, 89-93.

[14] Kobayashi, T.;Bando, M.;Kimura, N.;Shimizu, K.;Kadono, K.;Umezu, N.;Miyahara, K.;Hayazaki, S.;Nagai, S.;Mizuguchi, Y.;Murakami, Y.; Hobara, D. Production of a 100-m-long high-quality graphene transparent conductive film by roll-to-roll chemical vapor deposition and transfer process. *Applied Physics Letters* **2013**, *102*, 1-4.

[15] Andrea, L.;Konstantinos, K.-A.;Xavier Diez, B.;Alessandro, K.;Emanuele, T.;Nicola Maria, P.;Giovanna De, L.;Loris, G.; Vincenzo, P. Evolution of the size and shape of 2D nanosheets during ultrasonic fragmentation. *2D Materials* **2017**, *4*, 025017.

[16] Palermo, V.;Kinloch, I. A.;Ligi, S.; Pugno, N. M. Nanoscale Mechanics of Graphene and Graphene Oxide in Composites: A Scientific and Technological Perspective. *Adv Mater* **2016**, *28*, 6232-6238.

[17] Ago, H. CVD Growth of High-Quality Single-Layer Graphene. In *Frontiers of Graphene and Carbon Nanotubes: Devices and Applications*. K. Matsumoto, Ed.; Springer Japan; Tokyo, 2015; pp 3-20.

[18] Novoselov, K. S.;Falko, V. I.;Colombo, L.;Gellert, P. R.;Schwab, M. G.; Kim, K. A roadmap for graphene. *Nature* **2012**, *490*, 192-200.

[19] Anagnostopoulos, G.;Pappas, P.-N.;Li, Z.;Kinloch, I. A.;Young, R. J.;Novoselov, K. S.;Lu, C. Y.;Pugno, N.;Parthenios, J.;Galiotis, C.; Papagelis, K. Mechanical Stability of Flexible Graphene-Based Displays. *ACS Applied Materials & Interfaces* **2016**, *8*, 22605-22614.

[20] Zhu, W.;Low, T.;Perebeinos, V.;Bol, A. A.;Zhu, Y.;Yan, H.;Tersoff, J.; Avouris, P. Structure and Electronic Transport in Graphene Wrinkles. *Nano Lett* **2012**, *12*, 3431-3436.

[21] Kim, K.;Lee, Z.;Malone, B. D.;Chan, K. T.;Alemán, B.;Regan, W.;Gannett, W.;Crommie, M. F.;Cohen, M. L.; Zettl, A. Multiply folded graphene. *Phys Rev B* **2011**, *83*, 245433.

[22] Kim, K.;Artyukhov, V. I.;Regan, W.;Liu, Y.;Crommie, M. F.;Yakobson, B. I.; Zettl, A. Ripping Graphene: Preferred Directions. *Nano Lett* **2011**, *12*, 293-297.

[23] Wang, Y.;Yang, R.;Shi, Z.;Zhang, L.;Shi, D.;Wang, E.; Zhang, G. Super-Elastic Graphene Ripples for Flexible Strain Sensors. *Acs Nano* **2011**, *5*, 3645-3650.

[24] Androulidakis, C.;Koukaras, E. N.;Pastore Carbone, M. G.;Hadjinicolaou, M.; Galiotis, C. Wrinkling formation in simply-supported graphenes under tension and compression loadings. *Nanoscale* **2017**, 10.1039/C7NR06463B.

[25] Androulidakis, C.;Koukaras, E. N.;Rahova, J.;Sampathkumar, K.;Parthenios, J.;Papagelis, K.;Frank, O.; Galiotis, C. Wrinkled Few-Layer Graphene as Highly Efficient Load Bearer. *ACS Applied Materials & Interfaces* **2017**, *9*, 26593-26601.

[26] Li, L.;Reich, S.; Robertson, J. Defect energies of graphite: Density-functional calculations. *Phys Rev B* **2005**, *72*, 184109.

[27] Cortijo, A.; Vozmediano, M. A. H. Electronic properties of curved graphene sheets. *EPL (Europhysics Letters)* **2007**, *77*, 47002.

[28] Ariza, M. P.; Ortiz, M. Discrete dislocations in graphene. *Journal of the Mechanics and Physics of Solids* **2010**, *58*, 710-734.

[29] Duplock, E. J.;Scheffler, M.; Lindan, P. J. D. Hallmark of Perfect Graphene. *Phys Rev Lett* **2004**, *92*, 225502.

[30] Lherbier, A.;Blase, X.;Niquet, Y.-M.;Triozon, F.; Roche, S. Charge Transport in Chemically Doped 2D Graphene. *Phys Rev Lett* **2008**, *101*, 036808.

[31] Han, J.;Pugno, N. M.; Ryu, S. Nanoindentation cannot accurately predict the tensile strength of graphene or other 2D materials. *Nanoscale* **2015**, *7*, 15672-15679.

[32] Jiang, T.;Huang, R.; Zhu, Y. Interfacial Sliding and Buckling of Monolayer Graphene on a Stretchable Substrate. *Adv Funct Mater* **2014**, *24*, 396-402.

[33] Bronsgeest, M. S.;Bendiab, N.;Mathur, S.;Kimouche, A.;Johnson, H. T.;Coraux, J.; Pochet, P. Strain Relaxation in CVD Graphene: Wrinkling with Shear Lag. *Nano Lett* **2015**, *15*, 5098-5104.

[34] Li, Z.;Kinloch, I. A.;Young, R. J.;Novoselov, K. S.;Anagnostopoulos, G.;Parthenios, J.;Galiotis, C.;Papagelis, K.;Lu, C.-Y.; Britnell, L. Deformation of Wrinkled Graphene. *Acs Nano* **2015**, *9*, 3917-3925.

[35] Vasić, B.;Zurutuza, A.; Gajić, R. Spatial variation of wear and electrical properties across wrinkles in chemical vapour deposition graphene. *Carbon* **2016**, *102*, 304-310.

[36] Tong Y, B. S., Song M. Graphene based materials and their composites as coatings. *Austin Journal of Nanomedicine & Nanotechnology* **2014**, *1*, 1-16.

[37] Raju, A. P. A.;Lewis, A.;Derby, B.;Young, R. J.;Kinloch, I. A.;Zan, R.; Novoselov, K. S. Wide-Area Strain Sensors based upon Graphene-Polymer Composite Coatings Probed by Raman Spectroscopy. *Adv Funct Mater* **2014**, *24*, 2865-2874.

[38] Won, S.;Hwangbo, Y.;Lee, S.-K.;Kim, K.-S.;Kim, K.-S.;Lee, S.-M.;Lee, H.-J.;Ahn, J.-H.;Kim, J.-H.; Lee, S.-B. Double-layer CVD graphene as stretchable transparent electrodes. *Nanoscale* **2014**, *6*, 6057-6064.

[39] Lee, Y.;Bae, S.;Jang, H.;Jang, S.;Zhu, S.-E.;Sim, S. H.;Song, Y. I.;Hong, B. H.; Ahn, J.-H. Wafer-Scale Synthesis and Transfer of Graphene Films. *Nano Lett* **2010**, *10*, 490-493.

[40] Kim, Y.-J.;Cha, J. Y.;Ham, H.;Huh, H.;So, D.-S.; Kang, I. Preparation of piezoresistive nano smart hybrid material based on graphene. *Current Applied Physics* **2011**, *11*, S350-S352.

[41] Bonaccorso, F.;Sun, Z.;Hasan, T.; Ferrari, A. C. Graphene photonics and optoelectronics. *Nat Photon* **2010**, *4*, 611-622.

[42] Kim, K. S.;Zhao, Y.;Jang, H.;Lee, S. Y.;Kim, J. M.;Kim, K. S.;Ahn, J.-H.;Kim, P.;Choi, J.-Y.; Hong, B. H. Large-scale pattern growth of graphene films for stretchable transparent electrodes. *Nature* **2009**, *457*, 706-710.

[43] Kim, B. J.;Jang, H.;Lee, S.-K.;Hong, B. H.;Ahn, J.-H.; Cho, J. H. High-Performance Flexible Graphene Field Effect Transistors with Ion Gel Gate Dielectrics. *Nano Lett* **2010**, *10*, 3464-3466.

[44] Schrier, J. Helium Separation Using Porous Graphene Membranes. *The Journal of Physical Chemistry Letters* **2010**, *1*, 2284-2287.

[45] Banhart, F.;Kotakoski, J.; Krasheninnikov, A. V. Structural Defects in Graphene. *Acs Nano* **2010**, *5*, 26-41.

[46] Berger, D.; Ratsch, C. Line defects in graphene: How doping affects the electronic and mechanical properties. *Phys Rev B* **2016**, *93*, 235441.

[47] Viera Skakalova, A. B. K. *Graphene: Properties, Preparation, Characterisation and Devices*; Woodhead Publishing Ltd, 2014.

[48] Hu, K.;Kulkarni, D. D.;Choi, I.; Tsukruk, V. V. Graphene-polymer nanocomposites for structural and functional applications. *Progress in Polymer Science* **2014**, *39*, 1934-1972.

[49] Mittal, G.;Dhand, V.;Rhee, K. Y.;Park, S.-J.; Lee, W. R. A review on carbon nanotubes and graphene as fillers in reinforced polymer nanocomposites. *Journal of Industrial and Engineering Chemistry* **2015**, *21*, 11-25.

[50] Terrones, M.;Martín, O.;González, M.;Pozuelo, J.;Serrano, B.;Cabanelas, J. C.;Vega-Díaz, S. M.; Baselga, J. Interphases in Graphene Polymer-based Nanocomposites: Achievements and Challenges. *Adv Mater* **2011**, *23*, 5302-5310.

[51] Murakami, K.;Kadowaki, T.; Fujita, J.-i. Damage and strain in single-layer graphene induced by very-low-energy electron-beam irradiation. *Applied Physics Letters* **2013**, *102*, 043111.

[52] Imamura, G.; Saiki, K. UV-irradiation induced defect formation on graphene on metals. *Chemical Physics Letters* **2013**, *587*, 56-60.

[53] Lucchese, M. M.;Stavale, F.;Ferreira, E. H. M.;Vilani, C.;Moutinho, M. V. O.;Capaz, R. B.;Achete, C. A.; Jorio, A. Quantifying ion-induced defects and Raman relaxation length in graphene. *Carbon* **2010**, *48*, 1592-1597.

[54] Cançado, L. G.;Jorio, A.;Ferreira, E. H. M.;Stavale, F.;Achete, C. A.;Capaz, R. B.;Moutinho, M. V. O.;Lombardo, A.;Kulmala, T. S.; Ferrari, A. C. Quantifying Defects in Graphene via Raman Spectroscopy at Different Excitation Energies. *Nano Lett* **2011**, *11*, 3190-3196.

[55] Eckmann, A.;Felten, A.;Mishchenko, A.;Britnell, L.;Krupke, R.;Novoselov, K. S.; Casiraghi, C. Probing the Nature of Defects in Graphene by Raman Spectroscopy. *Nano Lett* **2012**, *12*, 3925-3930.

[56] Kim, B. H.;Hong, S. J.;Baek, S. J.;Jeong, H. Y.;Park, N.;Lee, M.;Lee, S. W.;Park, M.;Chu, S. W.;Shin, H. S.;Lim, J.;Lee, J. C.;Jun, Y.; Park, Y. W. N-type graphene induced by dissociative H2 adsorption at room temperature. *Scientific Reports* **2012**, *2*, 690.

[57] Allouche, A.; Ferro, Y. Dissociative adsorption of small molecules at vacancies on the graphite (0 0 0 1) surface. *Carbon* **2006**, *44*, 3320-3327.

[58] Diño, W. A.;Miura, Y.;Nakanishi, H.;Kasai, H.;Sugimoto, T.; Kondo, T. H2 dissociative adsorption at the armchair edges of graphite. *Solid State Communications* **2004**, *132*, 713-718.

[59] Anagnostopoulos, G.;Androulidakis, C.;Koukaras, E. N.;Tsoukleri, G.;Polyzos, I.;Parthenios, J.;Papagelis, K.; Galiotis, C. Stress Transfer Mechanisms at the Submicron Level for Graphene/Polymer Systems. *ACS Applied Materials & Interfaces* **2015**, *7*, 4216-4223.

[60] Jeong, H.-K.;Lee, Y. P.;Lahaye, R. J. W. E.;Park, M.-H.;An, K. H.;Kim, I. J.;Yang, C.-W.;Park, C. Y.;Ruoff, R. S.; Lee, Y. H. Evidence of Graphitic AB Stacking Order of Graphite Oxides. *Journal of the American Chemical Society* **2008**, *130*, 1362-1366.

[61] Koh, Y. K.;Bae, M.-H.;Cahill, D. G.; Pop, E. Reliably Counting Atomic Planes of Few-Layer Graphene (n > 4). *Acs Nano* **2010**, *5*, 269-274.

[62] Despiau-Pujo, E.;Davydova, A.;Cunge, G.; Graves, D. B. Hydrogen Plasmas Processing of Graphene Surfaces. *Plasma Chemistry and Plasma Processing* **2016**, *36*, 213-229.

[63] Wu, Z.-S.;Ren, W.;Gao, L.;Zhao, J.;Chen, Z.;Liu, B.;Tang, D.;Yu, B.;Jiang, C.; Cheng, H.-M. Synthesis of Graphene Sheets with High Electrical Conductivity and Good Thermal Stability by Hydrogen Arc Discharge Exfoliation. *Acs Nano* **2009**, *3*, 411-417.

[64] Ado Jorio, E. H. M. F., Luiz G. Cançado, Carlos A. Achete and Rodrigo B. Capaz Measuring Disorder in Graphene with Raman Spectroscopy, . In *Physics and Applications of Graphene - Experiments*. D. S. Mikhailov, Ed.; InTech, 2011.

[65] Siokou, A.;Ravani, F.;Karakalos, S.;Frank, O.;Kalbac, M.; Galiotis, C. Surface refinement and electronic properties of graphene layers grown on copper substrate: An XPS, UPS and EELS study. *Applied Surface Science* **2011**, *257*, 9785-9790.

[66] Luo, Z.;Shang, J.;Lim, S.;Li, D.;Xiong, Q.;Shen, Z.;Lin, J.; Yu, T. Modulating the electronic structures of graphene by controllable hydrogenation. *Applied Physics Letters* **2010**, *97*, 233111.

[67] Kim, J.-H.;Hwang, J. H.;Suh, J.;Tongay, S.;Kwon, S.;Hwang, C. C.;Wu, J.; Park, J. Y. Work function engineering of single layer graphene by irradiation-induced defects. *Applied Physics Letters* **2013**, *103*, 171604.

[68] Ilie, A.;Hart, A.;Flewitt, A. J.;Robertson, J.; Milne, W. I. Effect of work function and surface microstructure on field emission of tetrahedral amorphous carbon. *J Appl Phys* **2000**, *88*, 6002-6010.

[69] Ferrari, A. C.; Basko, D. M. Raman spectroscopy as a versatile tool for studying the properties of graphene. *Nat Nano* **2013**, *8*, 235-246.

[70] Liu, N.;Pan, Z.;Fu, L.;Zhang, C.;Dai, B.; Liu, Z. The origin of wrinkles on transferred graphene. *Nano Res* **2011**, *4*, 996-1004.

[71] Li, Y.; Chopra, N. Progress in Large-Scale Production of Graphene. Part 2: Vapor Methods. *JOM* **2014**, *67*, 44-52.

[72] Chen, P. Y.;Sodhi, J.;Qiu, Y.;Valentin, T. M.;Steinberg, R. S.;Wang, Z.;Hurt, R. H.; Wong, I. Y. Multiscale graphene topographies programmed by sequential mechanical deformation. *Advanced Materials* **2016**, *28*, 3564-3571.

[73] Jiang, T.; Zhu, Y. Measuring graphene adhesion using atomic force microscopy with a microsphere tip. *Nanoscale* **2015**, *7*, 10760-10766.

[74] Maugis, D. Adhesion of spheres: The JKR-DMT transition using a dugdale model. *Journal of Colloid and Interface Science* **1992**, *150*, 243-269.

[75] Shin, Y. J.;Wang, Y.;Huang, H.;Kalon, G.;Wee, A. T. S.;Shen, Z.;Bhatia, C. S.; Yang, H. Surface-Energy Engineering of Graphene. *Langmuir* **2010**, *26*, 3798-3802.

[76] Kozbial, A.;Li, Z.;Conaway, C.;McGinley, R.;Dhingra, S.;Vahdat, V.;Zhou, F.;D'Urso, B.;Liu, H.; Li, L. Study on the Surface Energy of Graphene by Contact Angle Measurements. *Langmuir* **2014**, *30*, 8598-8606.

[77] Malard, L. M.;Pimenta, M. A.;Dresselhaus, G.; Dresselhaus, M. S. Raman spectroscopy in graphene. *Physics Reports* **2009**, *473*, 51-87.

[78] Dresselhaus, M. S.;Jorio, A.;Hofmann, M.;Dresselhaus, G.; Saito, R. Perspectives on Carbon Nanotubes and Graphene Raman Spectroscopy. *Nano Lett* **2010**, *10*, 751-758.

[79] Berciaud, S.;Ryu, S.;Brus, L. E.; Heinz, T. F. Probing the Intrinsic Properties of Exfoliated Graphene: Raman Spectroscopy of Free-Standing Monolayers. *Nano Lett* **2009**, *9*, 346-352.

[80] Androulidakis, C.;Koukaras, E. N.;Parthenios, J.;Kalosakas, G.;Papagelis, K.; Galiotis, C. Graphene flakes under controlled biaxial deformation. *Scientific Reports* **2015**, *5*, 18219.

[81] Frank, O.;Vejpravova, J.;Holy, V.;Kavan, L.; Kalbac, M. Interaction between graphene and copper substrate: The role of lattice orientation. *Carbon* **2014**, *68*, 440-451.

[82] Wood, J. D.;Schmucker, S. W.;Lyons, A. S.;Pop, E.; Lyding, J. W. Effects of Polycrystalline Cu Substrate on Graphene Growth by Chemical Vapor Deposition. *Nano Lett* **2011**, *11*, 4547-4554.

[83] Ferrari, A. C.; Basko, D. M. (REVIEW) Raman spectroscopy as a versatile tool for studying the properties of graphene. *Nat Nano* **2013**, *8*, 235-246.

[84] He, R.;Zhao, L.;Petrone, N.;Kim, K. S.;Roth, M.;Hone, J.;Kim, P.;Pasupathy, A.; Pinczuk, A. Large Physisorption Strain in Chemical Vapor Deposition of Graphene on Copper Substrates. *Nano Lett* **2012**, *12*, 2408-2413.

[85] Yu, Q.;Jauregui, L. A.;Wu, W.;Colby, R.;Tian, J.;Su, Z.;Cao, H.;Liu, Z.;Pandey, D.;Wei, D.;Chung, T. F.;Peng, P.;Guisinger, N. P.;Stach, E. A.;Bao, J.;Pei, S.-S.; Chen, Y. P. Control and characterization of individual grains and grain boundaries in graphene grown by chemical vapour deposition. *Nat Mater* **2011**, *10*, 443-449.

[86] Reina, A.; Kong, J. Graphene Growth by CVD Methods. In *Graphene Nanoelectronics*. R. Murali, Ed.; Springer US, 2012; pp 167-203.

[87] Kalbac, M.;Frank, O.; Kavan, L. The control of graphene double-layer formation in copper-catalyzed chemical vapor deposition. *Carbon* **2012**, *50*, 3682-3687.

[88] Kabiri Ameri, S.;Ho, R.;Jang, H.;Tao, L.;Wang, Y.;Wang, L.;Schnyer, D. M.;Akinwande, D.; Lu, N. Graphene Electronic Tattoo Sensors. *Acs Nano* **2017**, *11*, 7634-7641.

[89] Zheng, J.;Wang, Y.;Wang, L.;Quhe, R.;Ni, Z.;Mei, W.-N.;Gao, Z.;Yu, D.;Shi, J.; Lu, J. Interfacial Properties of Bilayer and Trilayer Graphene on Metal Substrates. *Scientific Reports* **2013**, *3*, 2081.

[90] Kalbac, M.;Kong, J.; Dresselhaus, M. S. Raman Spectroscopy as a Tool to Address Individual Graphene Layers in Few-Layer Graphene. *The Journal of Physical Chemistry C* **2012**, *116*, 19046-19050.

[91] Martins Ferreira, E. H.;Moutinho, M. V. O.;Stavale, F.;Lucchese, M. M.;Capaz, R. B.;Achete, C. A.; Jorio, A. Evolution of the Raman spectra from single-, few-, and many-layer graphene with increasing disorder. *Phys Rev B* **2010**, *82*, 125429.

[92] Das, A.;Chakraborty, B.; Sood, A. K. Raman spectroscopy of graphene on different substrates and influence of defects. *Bull Mater Sci* **2008**, *31*, 579-584.

[93] Casiraghi, C. Probing disorder and charged impurities in graphene by Raman spectroscopy. *physica status solidi (RRL) – Rapid Research Letters* **2009**, *3*, 175-177.

[94] Ryan, B.;Luiz Gustavo, C.; Lukas, N. Raman characterization of defects and dopants in graphene. *Journal of Physics: Condensed Matter* **2015**, *27*, 083002.

[95] Guillermo López-Polín, C. G.-N., Vincenzo Parente, Francisco Guinea, Mikhail I. Katsnelson, Francesc Pérez-Murano, Julio Gómez-Herrero Stiffening graphene by controlled defect creation. *Materials Science* **2014**, 18.

[96] Papageorgiou, D. G.;Kinloch, I. A.; Young, R. J. Mechanical Properties of Graphene and Graphene-based Nanocomposites. *Progress in Materials Science* **2017**, *90*, 75-127.

[97] Mohiuddin, T. M. G.;Lombardo, A.;Nair, R. R.;Bonetti, A.;Savini, G.;Jalil, R.;Bonini, N.;Basko, D. M.;Galiotis, C.;Marzari, N.;Novoselov, K. S.;Geim, A. K.; Ferrari, A. C. Uniaxial strain in graphene by Raman spectroscopy: G peak splitting, Gruneisen parameters, and sample orientation. *Phys Rev B* **2009**, *79*, 205433.

[98] Tsoukleri, G.;Parthenios, J.;Papagelis, K.;Jalil, R.;Ferrari, A. C.;Geim, A. K.;Novoselov, K. S.; Galiotis, C. Subjecting a Graphene Monolayer to Tension and Compression. *Small* **2009**, *5*, 2397-2402.

[99] Tsoukleri, G.;Parthenios, J.;Galiotis, C.; Papagelis, K. Embedded trilayer graphene flakes under tensile and compressive loading. *2D Materials* **2015**, *2*, 024009.

[100] Bousa, M.;Anagnostopoulos, G.;del Corro, E.;Drogowska, K.;Pekarek, J.;Kavan, L.;Kalbac, M.;Parthenios, J.;Papagelis, K.;Galiotis, C.; Frank, O. Stress and charge transfer in uniaxially strained CVD graphene. *Physica Status Solidi (b)* **2016**, 10.1002/pssb.201600233, 1-7.

[101] Wang, G.;Dai, Z.;Liu, L.;Hu, H.;Dai, Q.; Zhang, Z. Tuning the Interfacial Mechanical Behaviors of Monolayer Graphene/PMMA Nanocomposites. *ACS Applied Materials & Interfaces* **2016**, *8*, 22554-22562.

[102] Frank, O.;Mohr, M.;Maultzsch, J.;Thomsen, C.;Riaz, I.;Jalil, R.;Novoselov, K. S.;Tsoukleri, G.;Parthenios, J.;Papagelis, K.;Kavan, L.; Galiotis, C. Raman 2D-Band Splitting in Graphene: Theory and Experiment. *Acs Nano* **2011**, *5*, 2231-2239.

[103]   Huang, M.;Yan, H.;Heinz, T. F.; Hone, J. Probing Strain-Induced Electronic Structure Change in Graphene by Raman Spectroscopy. *Nano Lett* **2010**, *10*, 4074-4079.

[104]   Zhang, R.;Ning, Z.;Zhang, Y.;Zheng, Q.;Chen, Q.;Xie, H.;Zhang, Q.;Qian, W.; Wei, F. Superlubricity in centimetres-long double-walled carbon nanotubes under ambient conditions. *Nat Nano* **2013**, *8*, 912-916.

[105]   Wang, G.;Liu, L.;Dai, Z.;Liu, Q.;Miao, H.; Zhang, Z. Biaxial compressive behavior of embedded monolayer graphene inside flexible poly (methyl methacrylate) matrix. *Carbon* **2015**, *86*, 69-77.

**TABLES**

**Table 1**: Ar$^+$ ion energy irradiation, number of impacts per cm$^2$, $sp^3/sp^2$ ratio, Work Function (WF) and I(D)/I(G) ratio of the corresponding graphene's Raman peaks for the pristine and treated CVD graphene/Cu surfaces.

| Ar$^+$ energy (eV) | Number of Ar$^+$ impacts/ cm$^2$ | $sp^3/sp^2$ | WF±0.05 (eV) | I(D)/I(G) |
|---|---|---|---|---|
| Pristine (0) | - | 0.10 | 4.40 | 0.13±0.03 |
| 35 | ~10$^{11}$ | 0.11 | 4.40 | 0.15 ± 0.05 |
| 65 | ~6x10$^{13}$ | 0.13 | 4.35 | 0.21 ± 0.09 |
| 100 | ~1x10$^{14}$ | 0.15 | 4.30 | 1.00 ± 0.32 |
| 120 | 1.5x10$^{14}$ | 0.17 | 4.30 | 3.36 ± 0.45 |
| 130 | 2x10$^{14}$ | 0.20 | 4.25 | 2.18 ± 0.54 |
| 200 | 6x10$^{14}$ | 0.23 | 4.20 | 0.96 ± 0.06 |

**Table 2**: The adhesion force and the surface energy between graphene and PMMA before and after the treatment with Ar$^+$

| Regions | Pristine | | Ar$^+$treated | |
|---|---|---|---|---|
| | *"Rugged"* | *"Flat"* | *"Rugged"* | *"Flat"* |
| **Adhesion force, *F* (nN)** | 2.8 | 3.2 | 5.84 | 5.18 |
| **Adhesion energy, *W*$_{adhesion}$ (mJ/m$^2$)** | 43.4 | 49.4 | 93.44 | 82.23 |

**Table 3**: The contact angle and the extracted surface energy values obtained by applying a water droplet on CVD graphene on Cu foil before and after the treatment with $Ar^+$

| Material | Contact angle (º) | Surface energy, $\gamma$ (mJ/m$^2$) |
|----------|-------------------|-------------------------------------|
| Pristine | 74.7 | 49.9 |
| Treated | 82.0 | 47.0 |

FIGURES

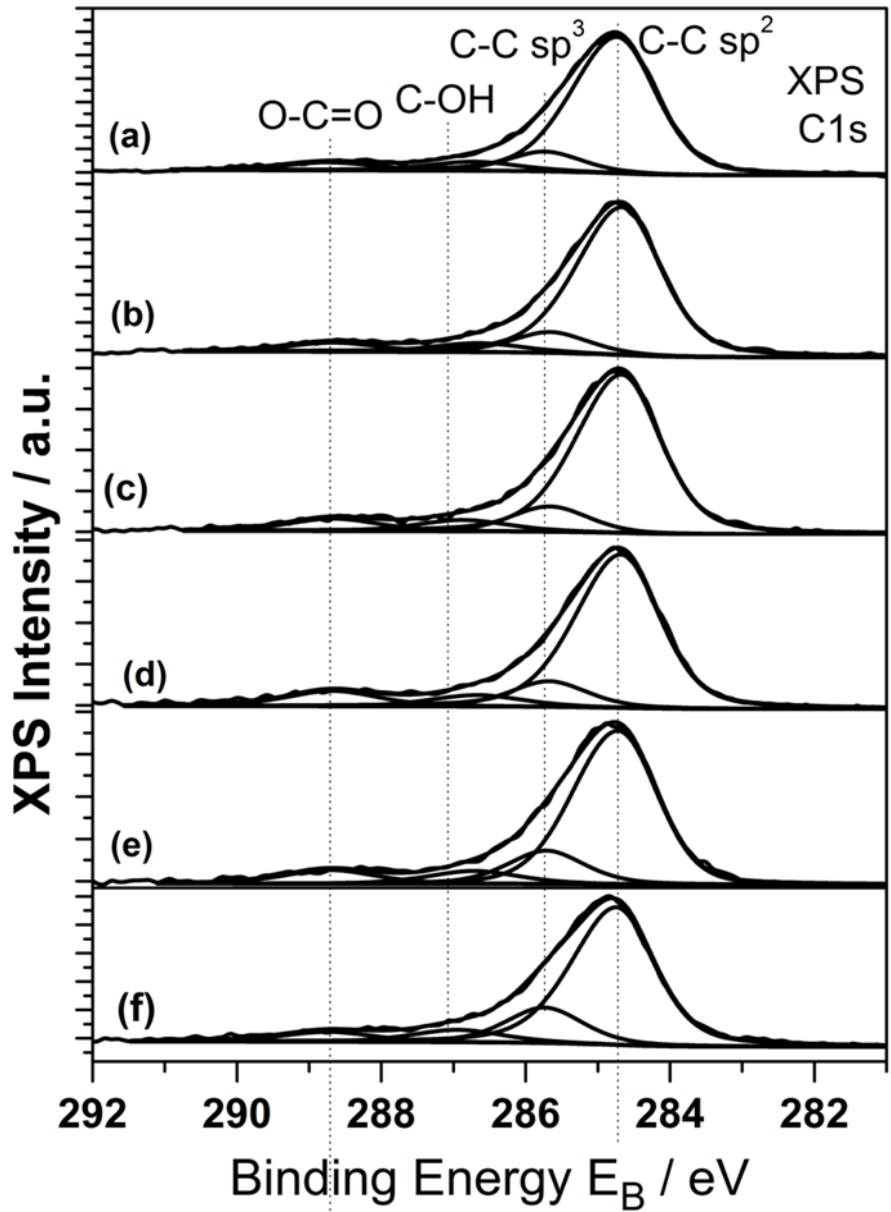

**Figure 1:** Deconvoluted C1s peak of CVD graphene/Cu samples for (a) pristine sample and after irradiation and exposed to $H_2$ for $Ar^+$ energies of (b) 35, (c) 65, (d) 100, (e) 130 and (f) 200 eV.

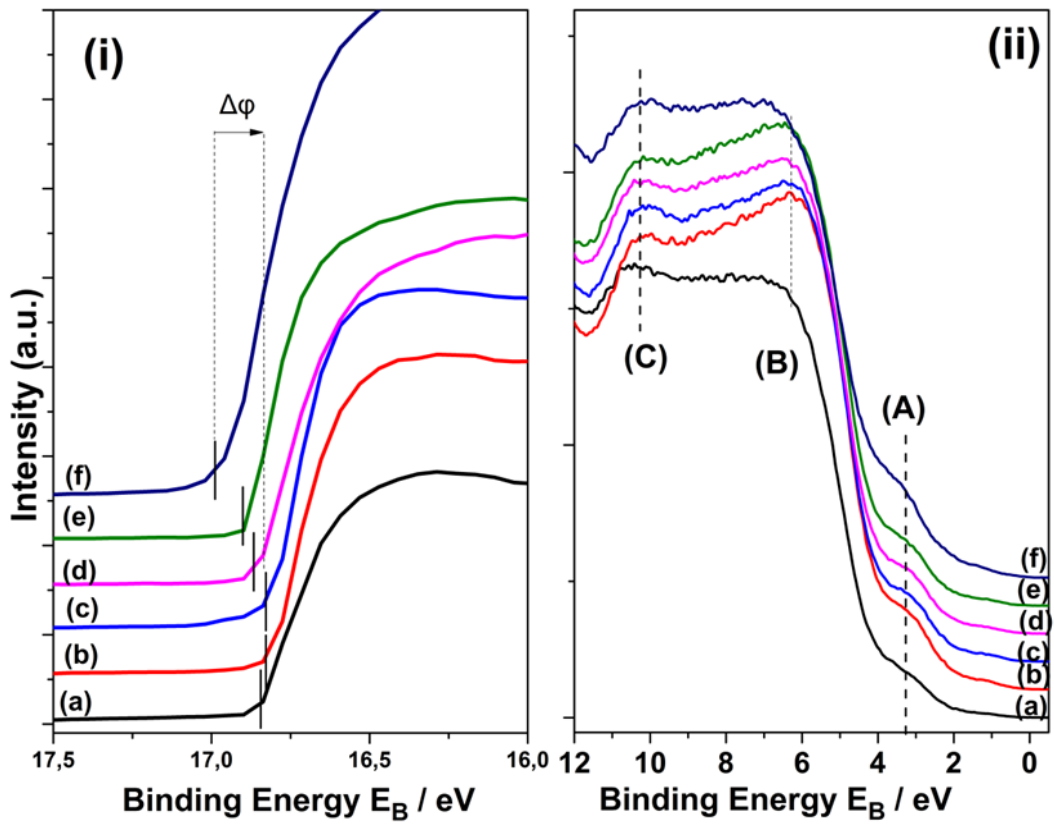

**Figure 2**: UPS spectra of CVD graphene/Cu samples. (I) valence band spectra and (II) the cut-off

for the (a) pristine sample and after irradiation and exposed to $H_2$ for $Ar^+$ energies of (b) 35, (c) 65,

(d) 100, (e) 130 and (f) 200 eV.

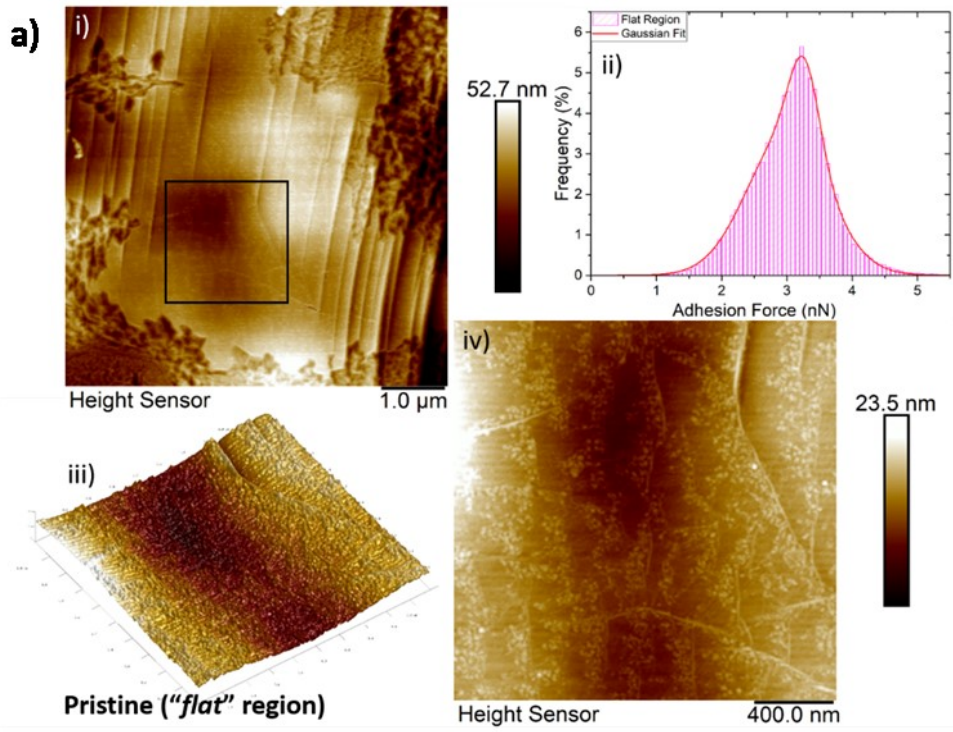

**a)** i) Height Sensor — 1.0 µm

52.7 nm

ii) Flat Region / Gaussian Fit

iii) **Pristine ("*flat*" region)**

iv) Height Sensor — 400.0 nm

23.5 nm

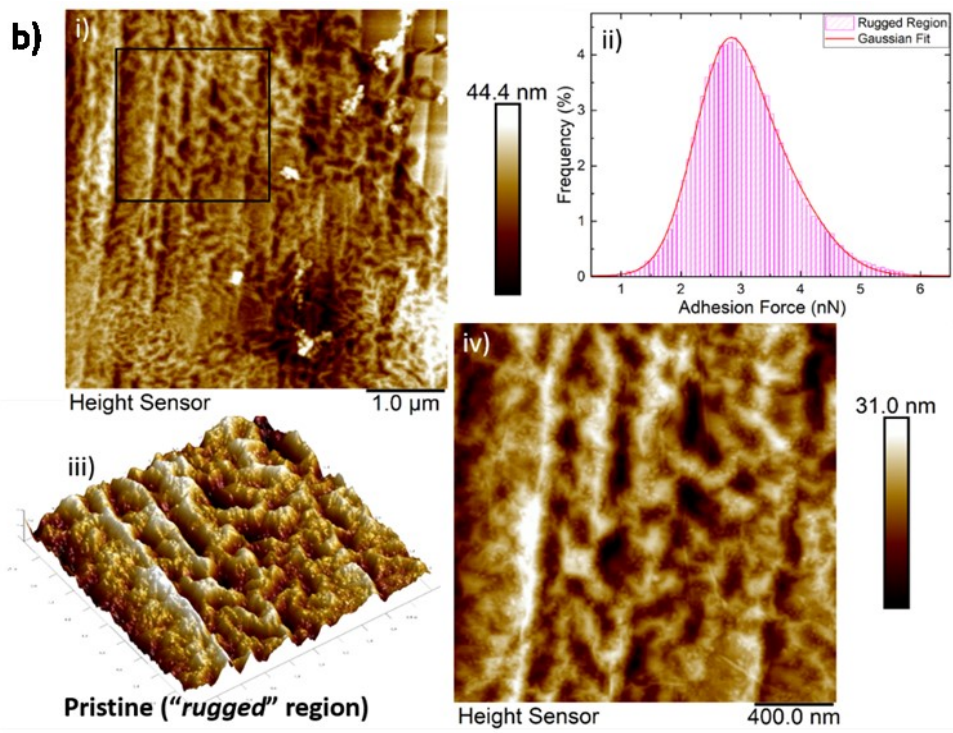

**b)** i) Height Sensor — 1.0 µm

44.4 nm

ii) Rugged Region / Gaussian Fit

iii) **Pristine ("*rugged*" region)**

iv) Height Sensor — 400.0 nm

31.0 nm

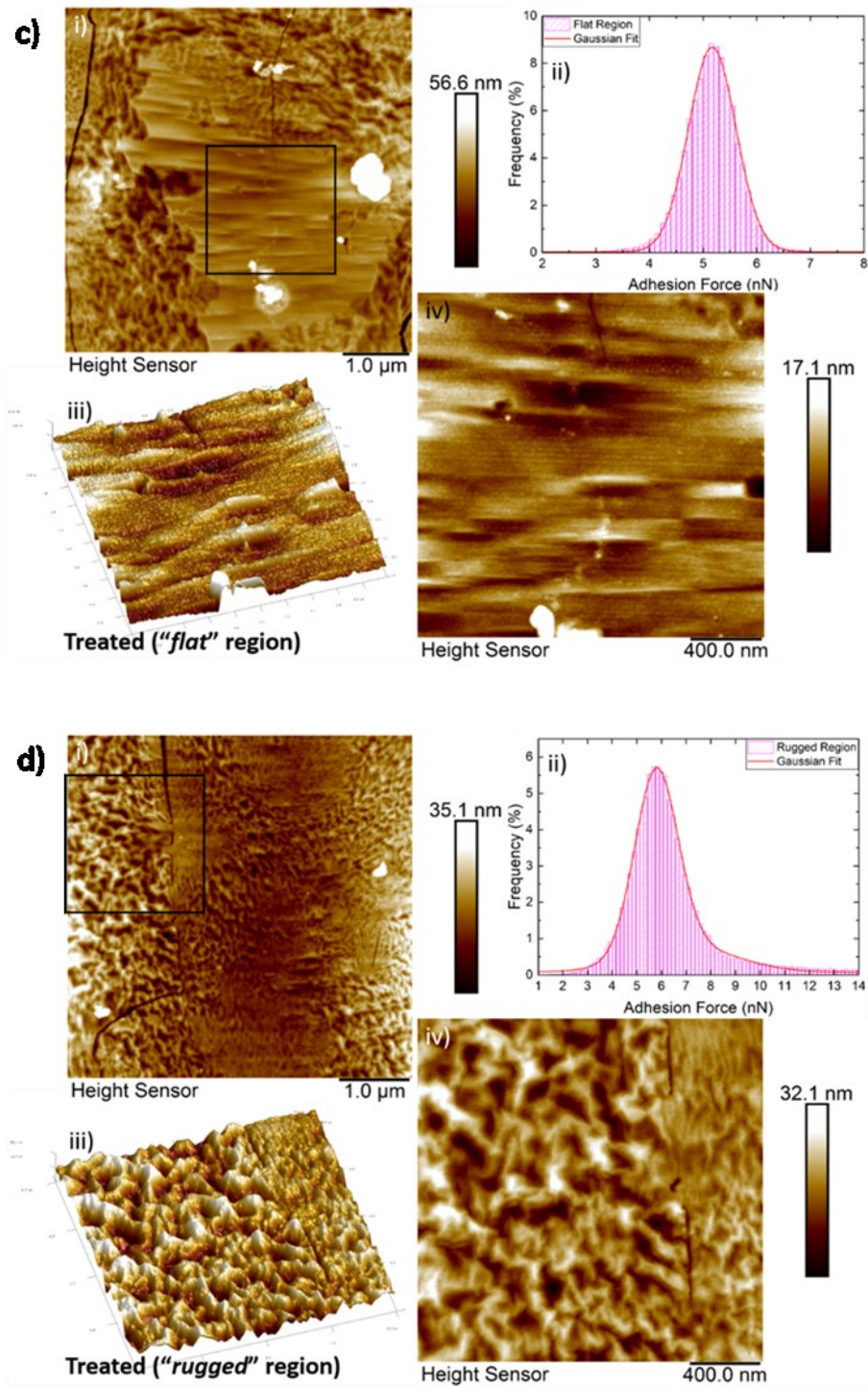

**Figure 3**: Atomic Force Microscopy (AFM) images prior loading of (a) "*flat*" region of pristine surface, (b) "*rugged*" region of pristine surface, (c) "*flat*" region of treated surface and (b) "*rugged*" region of treated surface. At every image, (i) represents the characteristic topography for each region while (iv) denotes the magnified region within the black solid square in (i). Also, (iii) is the corresponding 3D AFM image of (iv) and (ii) is the adhesion force histogram. All the samples

examined were irradiated with $Ar^+$ ion energy of 120 eV, which corresponds to $Ar^+$ impacts per $cm^2$ of $1.5x10^{14}Ar^+/cm^2$.

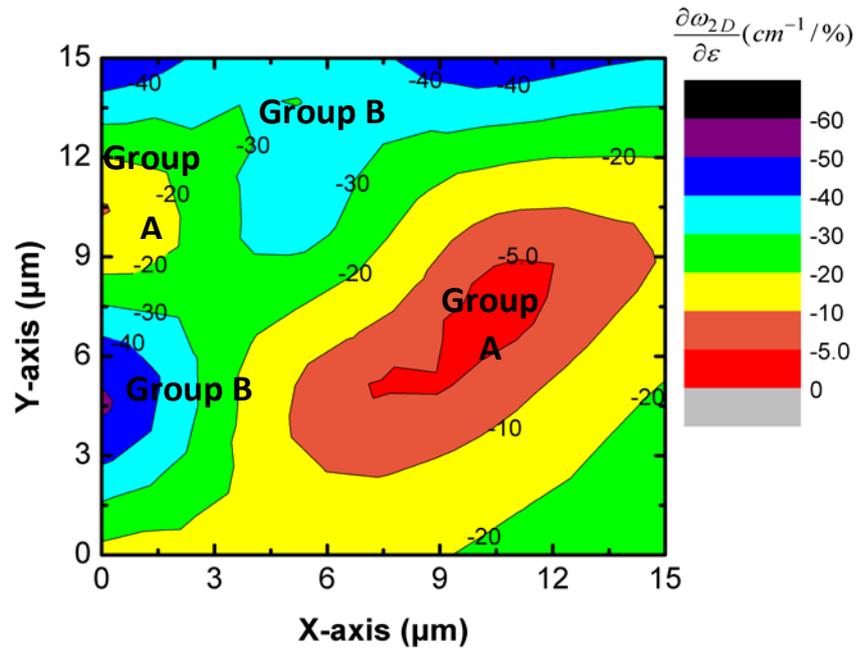

(a)

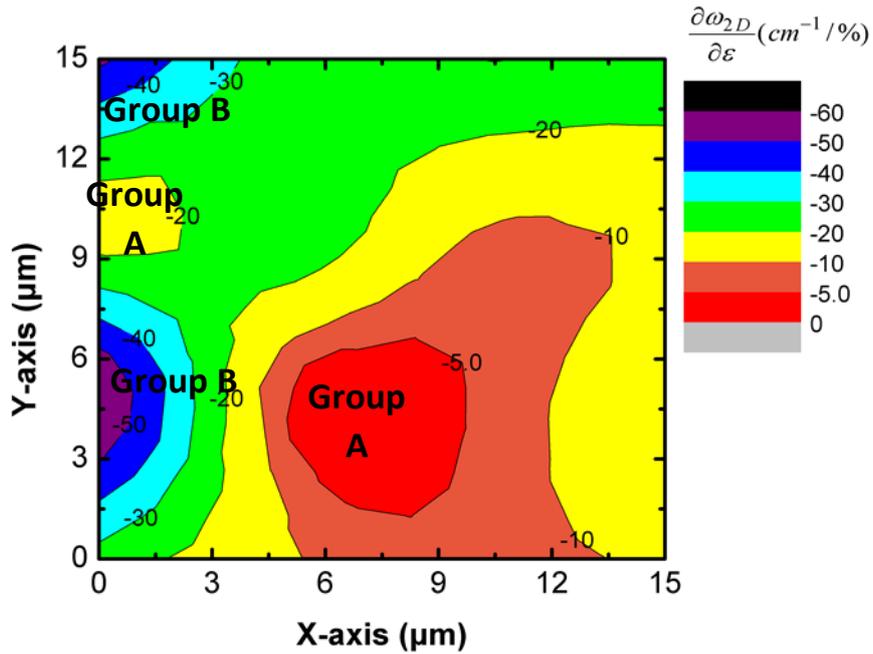

(b)

**Figure 4**: Strain rate maps for Pos(2D) of a 15 x15 µm²sub-area during (a) loading and (b) unloading for the pristine CVD graphene /PMMA system. Due to the apparent diversity of strain rate values of Pos(2D), two group of points with low (Group A) and high (Group B) strain rates are depicted.

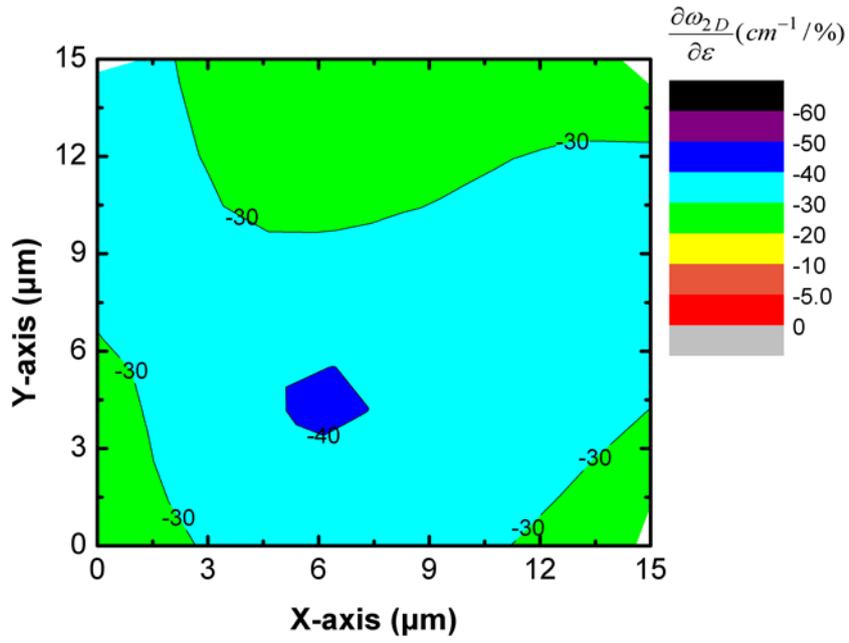

(a)

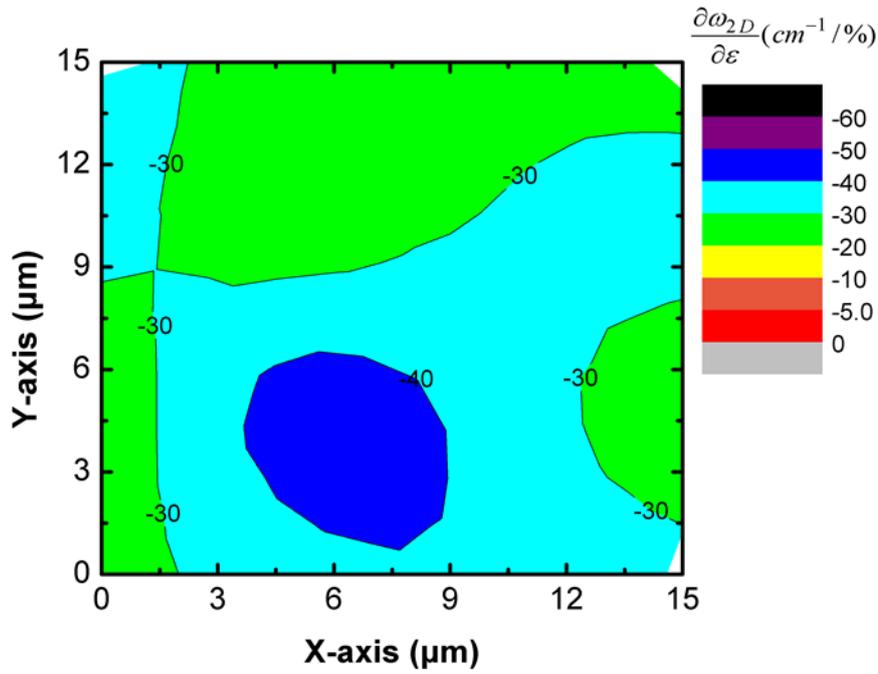

(b)

**Figure 5**: Strain rate maps for Pos(2D) of 15 x15 μm² area during (a) loading and (b) unloading for the treated CVD graphene /PMMA system, where an averaged increase of the shift over strain is observed in comparison to the pristine case. At the same time a very small diversity of strain rate values of Pos(2D) is also clearly depicted. The Ar⁺ ion energy irradiation took place at 120 eV, which corresponds to Ar⁺ impacts per cm² of 1.5x10¹⁴Ar⁺/cm².

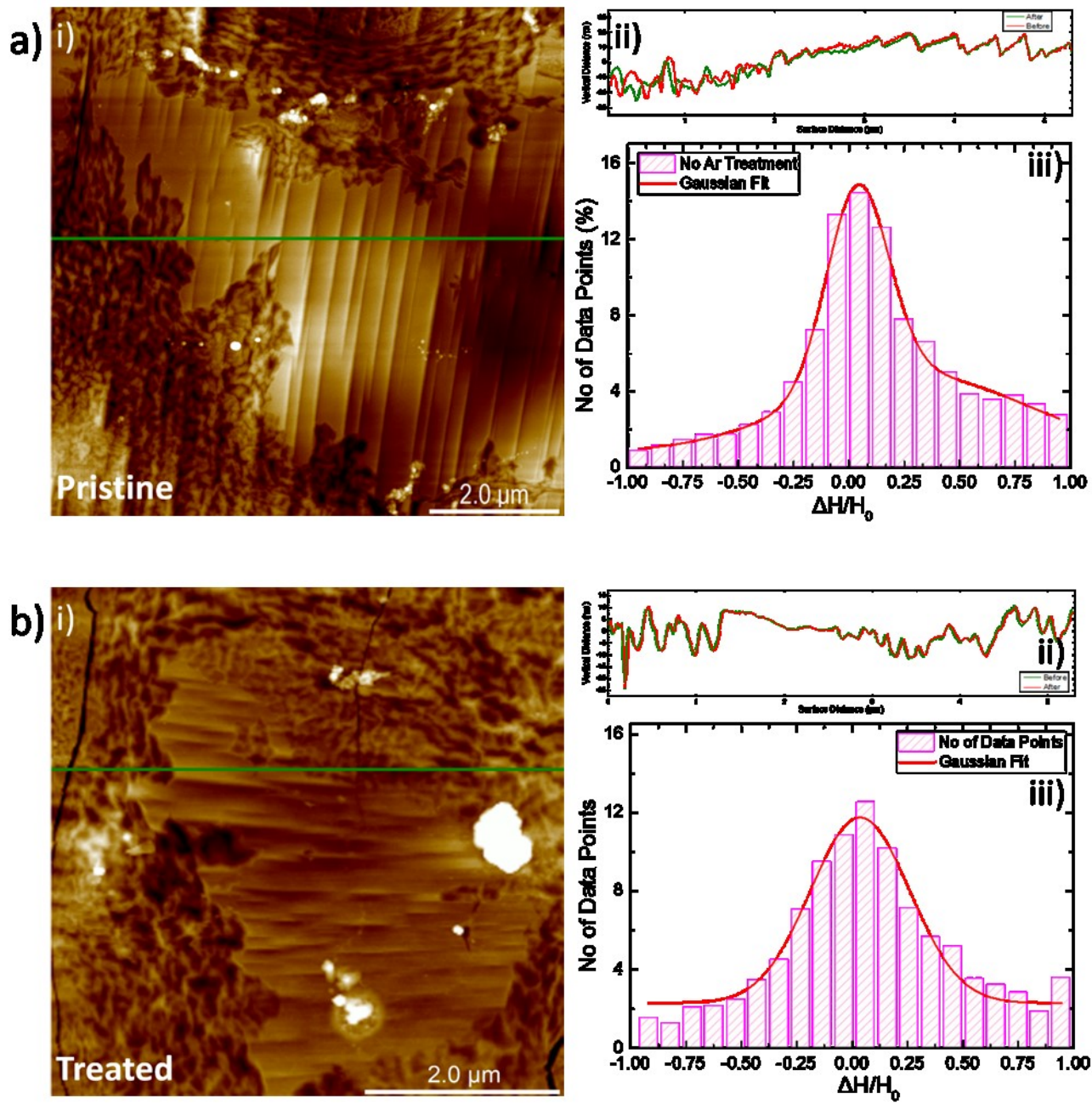

**Figure 6**: Atomic Force Microscopy (AFM) images of (a) pristine and (b) treated surface are showed in (i), while (ii) is the corresponding section analysis (green solid lines in (i) graphs) of the same region prior (green line) and after (red line) loading for the pristine and treated surface, respectively. In (iii) the statistical data of the relative change of surface height ($\Delta H/H_0$) are fitted with Gaussian distributions (red solid line) for both cases studied here.